\documentstyle[aps,prb,floats,epsf]{revtex}
\begin{document}
%
% for two column  activate the line below...
\twocolumn[\hsize\textwidth\columnwidth\hsize\csname @twocolumnfalse\endcsname
\title{Reformulation of the LDA+$U$ method for a local orbital basis}
\author{W.E. Pickett, S.C. Erwin, and E.C. Ethridge}
\address{Complex Systems Theory Branch, Naval Research Laboratory,
Washington DC 20375}
\date{\today}
\maketitle
\begin{abstract}
We present a new approach to the evaluation of the on-site repulsion
energy $U$ for use in the LDA+$U$ method of Anisimov and 
collaborators.  Our objectives are to make the method more firmly
based, to concentrate primarily on ground state properties rather
than spectra, and to test the method in cases where 
only modest changes in orbital occupations are expected, as well as for
highly correlated materials.
Because of these objectives, we employ a differential definition of $U$.
We also define a {\it matrix} ${\bf U}$, which we find is
very dependent on the environment of the atom in question.
The formulation is applied to evaluate $U$ for transition metal 
monoxides from VO to NiO
using a local orbital basis 
set.  The resulting values of $U$ are 
typically only 40-65\% as large
as values currently in use.  We evaluate the ${\bf U}$ matrix for
the $e_g$ 
and $t_{2g}$ subshells in paramagnetic FeO,
and illustrate the very different charge response
of the $e_g$ and $t_{2g}$ states.
The sensitivity of the method to the choice of the
$d$ orbitals, and to the basis set in general, is discussed.
\end{abstract}
\pacs{PACS numbers:This is a draft.}
%
% for two column  activate the line below...
]
\section{Introduction}

The understanding and evaluation of the electronic structure of strongly 
correlated materials is a long-standing problem.  For weakly correlated 
materials such as nearly-free-electron-like metals, covalent semiconductors, 
ionic solids, and even rather complex intermetallic transition metal 
compounds, the local density approximation (LDA, which we understand to 
include the spin degree of freedom as well) to the exchange-correlation 
functional that occurs in density functional theory gives very reasonable 
ground state properties and even band structures (which are excited state 
features).  For correlated materials, however,
LDA can be completely wrong: the now-classic example is the canonical cuprate
La$_2$CuO$_4$, which LDA predicts to be a non-magnetic metal
\cite{leung,weprmp} whereas it is 
actually an antiferromagnetic insulator.  Model many-body Hamiltonian 
treatments, such as the Hubbard model,\cite{hubbard}
can readily explain the observed type
of ground state but do so in terms of adjustable parameters and the neglect 
of many aspects of the crystal that may influence most of its properties.
Evaluation of the dynamic self-energy, which gives the description of
excitations, is appropriate for comparing with many experiments, but
even low-order approximations can be very tedious to evaluate.\cite{ferdi}

Within the past few years Anisimov and collaborators have proposed an 
extension of the LDA approach (now called LDA+$U$) based on lessons learned 
from Hubbard model studies\cite{hubbard} that single out a particular local orbital
and the associated on-site repulsive interaction $U$
as the fundamental characteristic to be 
addressed \cite{anigunn,aza,anietal,sasha}.  
They propose that the LDA treats the 
effects of $U$ reasonably well in some average sense, 
even in highly correlated systems, but that one must allow a deviation 
from this average behavior by including a correction
to the total energy including a term like 

\begin{eqnarray}
\Delta E = \frac{1}{2}\sum_{m,s\neq m',s'} \left( U-\delta_{s,s'}J \right) 
n_{ms} n_{m's'},
\label{schematic}
\end{eqnarray}
where $J$ is the exchange constant and $n_{ms}$ 
is the new charge that includes a local 
charge redistribution (relative to the LDA value $\bar n_{ms}$) 
obtained by solving the LDA+$U$ equations
self-consistently.  The local orbital and spin 
indices are $m$ and $s$,
respectively.  It is assumed in the method that
one can identify the orbitals to be treated ($d$~orbitals of Cu in 
La$_2$CuO$_4$ for the example mentioned above).

The LDA+$U$ method achieves some spectacular successes, such as leading to an 
antiferromagnetic insulating state of La$_2$CuO$_4$ with band gap and atomic moment in 
reasonable correspondence with observed values \cite{czyzyk}, and 
leading to similarly 
impressive descriptions of the transition metal monoxides.  There remain 
questions, however, such as the proper way to specify the orbitals, the
correct way to obtain the interaction constants ($U$ and $J$), and how, 
if possible, to extend the method to give an improved treatment of the metallic
phase when the insulator is heavily doped.  In this paper we address these
questions.  A primary feature is that, since the method is 
perforce focussed on an atomic orbital, it is natural to use a local orbital 
basis set.  We will refer to the local orbital of interest as the
``$d$ orbital(s)" although in some applications they may be $f$ or, rarely,
$s$ or $p$ orbitals.

\section{Description of LDA+$U$ as currently practiced}

In extending the LDA method to account for correlations resulting from
strong on-site interactions, there are several criteria that one might
hope to satisfy, such as (1) it should reduce to LDA when LDA is known
to be good; (2) the energy is given by a functional of the density;
(3) the method specifies how to obtain the local orbital in question
(perhaps through a self-consistency procedure); (4) the definition of
$U$ and $J$ are provided unambiguously; (5) the method predicts antiferromagnetic
insulators when appropriate; (6) the description of highly correlated
metals is improved over the LDA description.  This list, although perhaps
still incomplete, is already very ambitious, and only certain of these
desires have been addressed seriously.

Anisimov, Zaanen, and Andersen (AZA) \cite{aza} chose to refine the 
LDA by including an orbital-dependent
one-electron potential to explicitly account for the important Coulomb 
repulsions not treated fully in LDA. This was accomplished in analogy 
with Hartree-Fock theory by 
correcting the mean field contribution of the $d-d$ on-site interaction with an 
intra-atomic correction.  This correction has been applied in slightly
varying forms, but a representative example of the functional to be solved is
\begin{eqnarray}
E_{LDA+U} & = & E_{LDA} [n]-\frac{1}{2}U \sum_i N_i[{n_{ms}}]
(N_i[{n_{ms}}]-1) \nonumber \\ 
          &  &+ \frac{1}{2} \sum_i \sum_{ms \ne m's'} U _{m,m'}
                 n_{ims}n_{im's'}
\label{def of E}
\end{eqnarray}

Here $i$ denotes the lattice site, and terms involving $J$ have been 
neglected because we do not need to specify the complete form of the
functional for this paper.  $N_i$ is the site
sum of the $d$ charges, evaluated for the self-consistent LDA+$U$
densities.  The second term is presumed to be a 
reasonable description of the direct Coulomb interaction energy
included in the LDA expression.

Equation (2) reveals that, in the LDA+$U$ approach,
one singles out beforehand the atomic orbitals to be treated, and
decides how to specify them.  Implementations to date use
orbitals arising in the linearized muffin-tin orbital (LMTO) method.
The $d$ orbitals to which the $U$ correction is applied
are numerical solutions to a Schr\"odinger equation inside an
atomic sphere, and are zero outside this sphere.  In addition, the LDA+$U$ is
clearly no longer a straightforward density functional
because it depends on parameters $U$ and $J$ that depend on the LDA density
rather than the LDA+$U$ density.

The one-electron potential is the conventional LDA form of potential, plus an 
orbital-dependent shift of energy given by 
\begin{eqnarray}
\Delta V_{ms} & = U(\frac{1}{2}-n_{ms})
\label{def of V}
\end{eqnarray}
if $U_{mm'}  \rightarrow U$ is orbital independent.
The changes in the electronic structure are proportional to $U$, and
the definition and calculation of $U$ is the next topic to address.

To obtain $U$ and $J$, AZA performed LMTO calculations
for a supercell in which the $d$ charge on one atom is constrained
and the eigenvalue is obtained.\cite{dede}  
The $d$ orbitals on all atoms in the supercell are 
$\it decoupled$ entirely from the remaining part of the basis set.  This
makes the treatment of the local orbitals an ``atomic-like" problem,
which greatly reduces the difficulty associated with
constraining the occupation numbers.
It also has the effect of leaving a rather artificial system to
perform the screening.  For example, in NiO the screening system 
consists of oxygen p orbitals that cannot hybridize with the Ni
d~orbitals, plus whatever other virtual orbitals
are included in the basis set. 

The discreteness of the $d$ eigenvalues
makes it simple to specify the charge in
the spin-orbitals in the supercell, and $U$ and $J$
are determined from the relations
\begin{eqnarray}
U & = \varepsilon_{3d \uparrow} \left 
( \frac{\bar n}{2}+\frac{1}{2},\frac{\bar n}{2}\right) -  
\varepsilon_{3d \uparrow} \left ( \frac{\bar n}{2}+\frac{1}{2},
\frac{\bar n}{2}-1\right), 
\label{anisU}
\end{eqnarray}
in which the $d$ occupation differs by unity around a mean polarization
of unity, and
\begin{eqnarray}
J & = \varepsilon_{3d \uparrow} \left ( \frac{\bar n}{2}+\frac{1}{2},
\frac{\bar n}{2}-\frac{1}{2}\right) -
\varepsilon_{3d \downarrow} \left 
( \frac{\bar n}{2}+\frac{1}{2},\frac{\bar n}{2}-\frac{1}{2}\right),
\label{aniU}
\end{eqnarray}
which is a straightforward difference between up and down eigenvalues
for unit spin polarization.
Here $\varepsilon_{3d \uparrow}(n_{\uparrow},n_{\downarrow})$
( $\varepsilon_{3d \downarrow}(n_{\uparrow},n_{\downarrow})$) 
is the spin-up (spin-down) 3$d$ eigenvalue for occupancies $n_{\uparrow}$
and $n_{\downarrow}$. 

While it is widely recognized that the on-site repulsion $U$ is a screened 
quantity, the manner in which the screening should be done is not
precisely specified.  An early study by Cox, Coulthard, and Lloyd \cite{cox}
for 3$d$ metals used a renormalized neutral atom approach, although
it was recognized that screening processes might extend over a
somewhat larger region.  Anisimov and collaborators have chosen the method
presented in this section, but in this paper we pursue a different
approach, with a somewhat different objective.  
  
\section{Reformulation of LDA+$U$ for a local orbital basis}

We specify in following
subsections the various ways in which our approach differs from 
that in current use.

\subsection{LCAO Basis Set}

We begin with a basis set of local orbitals \{$\phi$\}, 
whose lattice sums lead to the 
standard linear combination of atomic orbitals (LCAO) Bloch basis functions 
for the one-electron Hamiltonian.\cite{erwin}  
To represent an occupied atomic orbital 
(including core states) we use a contracted set of 
Gaussian functions, multiplied 
by appropriate angular functions for $s$,~$p$,~or~$d$~behavior.  In particular, 
we choose at the beginning (from a neutral atom or an ion) the $d$~orbitals
 of central interest. Although we have no indication of any better choice 
than the $d$~orbital of the corresponding atom ({\it e.g.} neutral Cu 
in La$_2$CuO$_4$), our 
method allows the ability to check how sensitive the results are to the form 
chosen for the orbital.  In 
addition to basis functions describing filled atomic (or ionic) orbitals, we
add other Gaussian functions to the basis to provide a more nearly complete 
basis for the valence and conduction states than a minimal basis set 
would provide.  This feature is an advantage of our 
local orbital representation, 
as the ability to include self-consistent screening by a crystalline density of general form in the calculations 
is important.

This LCAO basis set brings up an important feature.  As a sum of squares of
wave functions, the charge density contains two types of terms.  One consists
of atom-centered contributions containing the coordinate dependence
$\phi^2_{l,m}(r-R)$ and is clearly identified as a contribution to the charge
from angular momentum $l$ of the atom located at R.  The other contribution
has a coordinate dependence given by $\phi_{lm}(r-R) \phi_{l'm'}(r-R')$, 
$R \neq R'$, which 
at a particular point may be positive or negative and cannot be assigned
uniquely to any atom.  The Mulliken decomposition,\cite{mulliken}
which assigns half of
each such term to charge component $l,m$ on the atom at R and the other half
to charge $l'm'$ of the atom at R', is widely used when atomic decomposition
of the charge is desired.  Mulliken population is well understood to be
not only arbitrary, but also dependent on the flexibility of the basis
set, and therefore should not be endowed with any important physical
meaning.  

A central fact that must be addressed is that the total
charge density cannot be decomposed, precisely or meaningfully, 
into simple atomic
contributions alone.  This fact means that the orbital occupations 
$n_{l,m}$ that are the centerpiece of the LDA+$U$ approach unfortunately are not
particularly well defined.  For our LCAO basis set we will use for 
$n_{ms}$ charge contributions solely
of the first type, which will be called on-site charges to
distinguish them from Mulliken charges.  These on-site quantities also cannot
properly be called occupation numbers since there is no sum rule for
their total, and it is not impossible that for a given orbital 
the value can exceed unity. 

\subsection{Specification of the functional}

Although we do not carry out LDA+$U$ calculations in this paper, we
are thinking in terms of a
generalized LDA+$U$ functional that is consistent with our philosophy
behind the correction.  Without more formal justification than is normally
done in the LDA+$U$ approach (and which we do not address seriously here), 
any change must simply be tested to see
if it produces better results.
The form
that we envision has affected our study of how to define and to evaluate
the interaction constants that arise in the method.  We 
suppose that the correction is to provide adjustment
to full potential LDA results, and therefore includes both a suborbital index
and a spin index on the reference charges $\bar n \rightarrow
\bar n_{ms}$.  These numbers will differ, sometimes greatly, for different
irreducible representations of the point group of the atom.  The
correction then might be written suggestively as 

\begin{eqnarray}
E_{{\rm LDA}+U}&=&E_{\rm LDA}[n]+\frac{1}{2} \sum_i \sum_{ms \neq m's'}(U _{m,m'}-
 \delta_{s,s'}J_{m,m'})    \nonumber  \\
  &   & \times (n_{ims}-\bar n_{ims})(n_{im's'}-\bar n_{im's'}).
\label{def of Ee}
\end{eqnarray}
This change may affect the types of orbitally-ordered solutions that
will be obtained.  This form ensures that the LDA solution is an
exact stationary solution of the LDA+$U$ functional (for which the correction
vanishes identically), which is not the case for Eq. (2), {\it i.e.}
if shell-averaged values
of $\bar n$ are used.  Aside from strongly correlated solids, another 
interest of ours 
is to ascertain whether LDA+$U$ can provide a useful improvement of the
description of `simpler' systems such as the transition metals Fe
and V, where anisotropy (relative amounts of t$_{2g}$
and e$_g$ character) is not reproduced accurately in
LDA, or in correlated metals where no bandgap occurs but charge
rearrangement might be appropriate.

\subsection{Procedure for determining $U$ and $J$}

We take as our ansatz that the constant $U$ (resp. $J$) occurring in the LDA+$U$
functional should describe the cost in potential energy of charge (resp. spin) 
fluctuations in the actual crystal, i.e. with all normal interactions 
and degrees of
freedom available
to the electrons.  Thus we do {\it not} decouple the $d$~states from surrounding
states. The $U$ and $J$ terms are going to be applied in precisely the same system 
from which they are determined.  We comment below on the question of 
dealing with the associated cost in kinetic energy due to charge fluctuations.
For the remainder of this paper we concentrate solely on $U$, postponing
a related treatment of $J$ such as suggested by Solovyev {\it et al.}
\cite{solovyev} for the future.

We also take the point of view that the main purpose of LDA+$U$ theory, as in 
density functional theory, is to obtain ground state properties, rather than
to approximate excitations with the eigenvalues.  
Describing the ground state may
require small rearrangements of occupation numbers away from their
LDA values, and usually less
than one-half, so we employ a differential definition of $U$ (also used
by Solovyev, Dederichs, and Anisimov \cite{solovyev}) rather than
one employing occupation numbers differing by unity.  We will see that
this introduces extra richness into the charge rearrangements described
by the LDA+$U$ method, because a small change in (say) $t_{2g}$ population can
be strongly compensated by a change in $e_g$ population.

We employ then a generalized constrained density functional approach as proposed
by Dederichs {\it et.al} \cite{dede} to calculate the 
change in energy due to constraints 
on local orbital densities.  We minimize the local
density functional subject to the constraint that on-site local orbital
charges 
{$n_{\alpha,s}$} be equal to designated values {$Q_{\alpha,s}$}, where $\alpha$
labels an irreducible representation of the point group of the atom in 
question ({\it e.g.} $t_{2g}$ or $e_g$), and that total charge $N$ be conserved:   
\begin{eqnarray}
{\cal E}({Q}) & = & \min_{n_{\uparrow},n_{\downarrow},{n_{\alpha,s}}} 
           \lbrace E_{\rm LDA}[n_{\uparrow},n_{\downarrow}] \nonumber \\
 &  & + \sum_{\alpha,s} w_{\alpha,s} (n_{\alpha,s} -Q_{\alpha,s}) \nonumber \\
 &  & -\mu (\int n(r)d^3r - N) \rbrace. 
\label{eq:minimize}
\end{eqnarray}
The Lagrange multipliers are the usual chemical potential $\mu$ and 
the potential shifts $w_{\alpha,s}$ necessary to satisfy the constraint
$n_{\alpha,s}=Q_{\alpha,s}$.  
Dependence on the total number $N$ of electrons 
(always conserved) will not be displayed explicitly.  
Variation with respect to the spin-orbitals
leads to a one-electron Schr\"odinger equation in which the potential is the LDA
potential, supplemented by local orbital shifts $w_{\alpha,s}$ on
the orbitals in the irreducible representation $\alpha$ 
having spin $s$.  These additional
shifts of potential can be represented as a non-local potential
\begin{eqnarray}
V_{NL} = \sum_{\alpha,s} \sum_{m \in \alpha} \mid \phi_{m,s}\rangle
w_{\alpha,s} \langle \phi_{m,s} \mid,
\label{eq:vnonloc}
\end{eqnarray}
where $\{\phi\}$ are normalized atomic orbitals.

Evaluation of the constrained energy in Eq. (7) deserves comment.  
Solution for the constrained energy involves generating the 
Kohn-Sham equations, which have an additional potential of the form
of Eq. (8) that effectively constrains the density as desired.  The
conventional method of evaluating the energy is to sum the resulting
eigenvalues and correct for double counting of the Hartree energy and
the miscounting of exchange-correlation energy.  That cannot be done
directly, because the Kohn-Sham eigenvalues contain the effects of
the additional potential of Eq. (8) and one does not obtain E$_{\rm LDA}$. 
The additional term that has been included by summing the LDA+$U$
eigenvalues however contains only the additional one-body term
$\sum_{\alpha,s} w_{\alpha,s} n_{\alpha,s}$, and this term can be
subtracted to obtain $E_{\rm LDA}$ {\it evaluated for the constrained
density}.

\subsection{The Constrained Energy}

It is convenient to introduce a vector notation for the local occupations,
the constraining values, and the (Lagrange parameter) potential shifts:
$n_{\alpha,s} \rightarrow \vec n$, and similarly for $\vec Q$ and for
$\vec w$.  Since we will be dealing with quantities relative to their
LDA values, we also use the notational conveniences
\begin{eqnarray}
\vec q & = & \vec Q - \vec Q^{\rm LDA} \nonumber \\
{\cal E}_{\vec q} & = & {\cal E}(\vec Q) - {\cal E}(\vec Q^{\rm LDA}). \nonumber 
\end{eqnarray}
From the Hellman-Feynman-like relation
\begin{eqnarray}
\frac {\partial {\cal E}_{\vec q}}{\partial \vec q} \equiv \nabla_{\vec q}
  {\cal E}_{\vec q} = - \vec w
\label{eq:partial}
\end{eqnarray} 
we can generalize the constrained density functional theory viewpoint of
Dederichs {\it et al.}\cite{dede} to obtain the change in 
energy due to constraining
a set of orbital densities in the manner of Eq. (7).  
Since there is no change in energy
if the charges are ``constrained" to be their LDA values $\vec Q =\vec
Q^{\rm LDA}$
(so $\vec q$ = 0),
the energy change is given by
\begin{eqnarray}
{\cal E}_{\vec q} = \int_{\vec 0}^{\vec q} d\vec q \cdot
     \nabla_{\vec q} {\cal E}_{\vec q} = -\int_{\vec 0}^{\vec q} d\vec q 
      \cdot \vec w(\vec q),
\label{eq:deltaE}
\end{eqnarray}
subject only to the condition that ${\cal E}_{\vec q}$ is analytic 
(as we assume).

The general behavior of the constrained energy can be seen by noting
that $\vec w$ is linear for small changes in
occupation, {\it i.e.} linear in $\vec q$. 
Since at the minimum of Eq. (7) $\vec n \equiv \vec Q$,
we may use these quantities interchangeably to write
\begin{eqnarray}
\vec w =- {\bf U} \vec q + {\cal O}(\vec q)^2
       =- {\bf U} \delta \vec n + 
   {\cal O}(\delta \vec n)^2,
\label{eq:wvsQ}
\end{eqnarray}
where $\delta \vec n$=$\vec n - \vec n^{\rm LDA}$ and ${\bf U}$ is the
constant (matrix) of proportionality.
For the remainder of this section we concern ourselves with the linear
``response" that is implicit in the LDA+$U$ method, although we 
demonstrate in the numerical results of Sec. V.A where non-linear
corrections begin to arise.
Then the energy shift is given by
\begin{eqnarray}
{\cal E}_{\vec q} = \frac {1}{2} \vec q \cdot {\bf U}
                 \cdot \vec q,
\end{eqnarray}
where 
\begin{eqnarray}
{\bf U} \equiv - \partial \vec w/\partial \vec q.
\end{eqnarray} 

The constrained energy ${\cal E}$($\vec q$) can be decomposed into the
kinetic energy term, the interaction with the external potential,
and the remainder, the potential energy:
\begin{eqnarray}
{\cal E}_{\vec q}~=~ {\cal E}_{\vec q,K}+
{\cal E}_{\vec q,{\rm ext}}+ {\cal E}_{\vec q,P}.
\end{eqnarray}
${\cal E}_{\vec q,{\rm ext}}$ is linear in $\vec q$ and gives no contribution to 
${\bf U}$,
but the quadratic term involving ${\bf U}$ contains both a kinetic 
energy contribution ${\bf U}_{\rm K}$ and a potential energy
contribution ${\bf U}_{\rm P}$,
\begin{eqnarray}
{\bf U}~=~{\bf U}_{\rm K}~+~{\bf U}_{\rm P}, \nonumber \\
{\bf U}_{\rm K(P)}~=~\nabla_{\vec q}\nabla_{\vec q} 
{\cal E}_{\vec q,{\rm K(P)}}.
\end{eqnarray}

In a self-consistent calculation, any change in local orbital charge results
in an accompanying change in kinetic energy as well as a potential energy
change.  Thus one might argue that
it should be ${\bf U}_{\rm P}$ that goes into the LDA+$U$ calculation, and the kinetic energy
change in the constrained LDA calculation should be removed: 
${\bf U} \rightarrow {\bf U}_{\rm P} = {\bf U} - {\bf U}_{\rm K}$ is the appropriate ``$U$'' in 
LDA+$U$.  Apparently this is
the correction (in our language) that Anisimov and Gunnarsson 
\cite{anigunn} expected to account for by disconnecting their local
orbitals from all other basis functions. 

Using the Hellman-Feynman relation Eq.(9) to obtain $\vec w(\vec q)$ 
it is straightforward to obtain ${\bf U}$ of Eq.(15). 
The kinetic energy contribution ${\bf U}_{\rm K}$ can be evaluated directly,
which is a very delicate task numerically, or alternatively from
the relation ${\cal E}_K=-[d{\cal E}/d\log m]_{m=m_o}$ which makes
use of the variational nature of the total energy ${\cal E}$ ($m_o$
is the electron mass).  Results for $U_K$ will be presented
elsewhere. 
For now, the value of $U$ that we evaluate and report below is 
the total value ${\bf U}$=${\bf U}_{\rm K}$+${\bf U}_{\rm P}$.

\subsection{Change in Independent Variable}

It will be instructive to consider the potential shifts $\vec w$ to
be the independent variables in an associated energy functional
leading to $\vec q(\vec w)$
rather 
than $\vec w(\vec q)$.
This is also in keeping with the practice in the constrained density
approach of choosing the shifts
$\vec w$ and then calculating the charge response
$\vec q$.   This change of variable is done by a Legendre transformation
\begin{eqnarray}
\hat {\cal E}_{\vec w} = {\cal E}_{\vec q} + \vec q \cdot \vec w,
\end{eqnarray}
which from the differential forms
\begin{eqnarray}
   \delta {\cal E}_{\vec q} = -\vec w \cdot \delta \vec q \nonumber \\
 \Rightarrow ~\delta \hat {\cal E}_{\vec w} = ~\vec q \cdot \delta \vec w,
\end{eqnarray}
leads to the energy shift
\begin{eqnarray}
\hat{\cal E}_{\vec w} =  ~\int_{\vec 0}^{\vec w} \vec q(\vec w)
      \cdot d \vec w \nonumber \\
  \approx - \frac {1}{2} \vec w \cdot {\bf U}^{-1} \cdot \vec w.
\end{eqnarray}

This formalism brings in the {\it matrix} ${\bf U}^{-1}$ implicit in Eq. (11)
relating the charge shifts
in various suborbitals to potential shifts applied to other suborbitals,
{\it e.g.} a decomposition of the Hubbard $U$ for $d$ orbitals into 
$e_g$ and $t_{2g}$ contributions for cubic site symmetry.  This result
is reminescent of the extension of of the definitions of $U$ and $J$ 
[Eqs. (4)-(5)] by Solovyev, Hamada, and
Terakura\cite{solovyev2} to give different values $U_{e_g}$ and $U_{t_{2g}}$,
but their procedure did not provide off-diagonal terms.  The effects
of differing charge response in the $e_g$ and $t_{2g}$ channels will
be quantified in Sec. V.  The concept can be extended to non-site-diagonal
interactions, {\it viz.} $d$ orbitals interacting with neighboring
oxygen $p$ orbitals.

We now establish a sum rule relating the matrix elements of ${\bf U}$ to
the conventional scalar $U$, which for clarity we denote $U_{dd}$
= $\partial w_d$/$\partial Q_d$, where Q$_d$ is the total $d$ charge
and $w_d$ is a shift in potential applied to all $d$ orbitals.
Since a change in potential $w_{t_{2g}}$ acting
on the $t_{2g}$ orbitals followed
by a change in potential $w_{e_g}$ acting on only the $e_g$ orbitals
is equivalent to a potential $w_d$ of the same magnitude acting on
all $d$ orbitals, we have, in the linear regime
\begin{eqnarray}
\frac{\partial ~}{\partial w_{t_{2g}}}~+~\frac{\partial~}{\partial
w_{e_g}}~=~\frac{\partial~}{\partial w_d}.
\end{eqnarray}
By definition $n_d=n_{t_{2g}}+n_{e_g}$, so from the definition 
\begin{eqnarray}
U_{\alpha \beta}^{-1}~=~-~\frac{\partial n_{\alpha}}{\partial w_{\beta}}
\end{eqnarray}
we have a sum rule relating the matrix elements to the conventional
Coulomb repulsion constant 
\begin{eqnarray}
U_{dd}^{-1}~=~\sum_{\alpha,\beta=t_{2g},e_g}
  U_{\alpha \beta}^{-1}.
\end{eqnarray}
We provide below a numerical test of this sum rule for NiO.

\section{Method of Calculation}

For the metallic constituents of the compounds we considered, a
basis set representing six {\it s-}, four {\it p-}, and three {\it d-}type
functions is expanded on a set of sixteen Gaussian functions.  
The O basis set is expanded on a set of twelve Gaussian exponents 
contracted into four {\it s-} and three {\it p-}type functions.
The Coulomb and exchange-correlation potentials comprise the effective 
potential, V$_{\rm eff}$, which is also described by a superposition of 
atom-centered Gaussian-type
functions. By choosing this expansion, the matrix elements of the Hamiltonian
are analytic.  Details of the method, and comparison to results of the
full potential linearized augmented plane wave method, have been
published elsewhere.\cite{erwin,erwin2} 

For this work it is important to obtain sufficiently well converged
values of orbital densities.  Tests using special point meshes
in the irreducible 1/48 of the simple cubic Brillouin 
zone (IBZ) for eight atom cells
up to 56 $\vec k$ points indicated that ten or twenty 
$\vec k$-points in the IBZ gave the necessary accuracy.
A temperature broadening of 0.07 eV was used to facilitate convergence
to self-consistency, and it was verified that this size of
broadening did not change the results.

\section{Evaluation for Transition Metal Monoxides} 

We have applied this approach to evaluate $U$ for the transitions metal
monoxides ${\it M}$O, ${\it M}$=V, Mn, Fe, Co, and Ni, in the 
paramagnetic state and for the cubic rocksalt structure.  The 
(experimental) lattice
constants used were: VO, 4.093 \AA; MnO, 4.444 \AA; FeO, 4.332
\AA; CoO, 4.260 \AA; NiO, 4.193 \AA.

\subsection{Sub-orbital Independent $U$}

\begin{table}[tbp]
\caption{Calculated values of $U$ for transition metal oxides, compared
to values of Anisimov, Zaanen, and Andersen (AZA).\protect\cite{aza}  Empirical values
include representative values from the literature.}
\begin{tabular}{ccccccc}
     Ref.  & VO  &  MnO & FeO & CoO & NiO \\
\tableline
This work     & 2.7  &  3.6  & 4.6  & 5.0  & 5.1  \\
 AZA          & 6.7  &  6.9  & 6.8  & 7.8  & 8.0  \\
Empirical &4.0-4.8$^a$&7.8-8.8$^a$&3.5-5.1$^a$&4.9-5.3$^a$&6.1-6.7$^a$ \\
          &           & 7.0$^d$ & 3.9$^c$,7.0$^d$& 4.9$^c$ & 7.9$^b$,
                                                       6.1$^c$, 7.5$^d$
\end{tabular}
\label{tableI}
$^a$V. I. Anisimov {\it et al.}, Ref. 6. \\
$^b$M. R. Norman and A. J. Freeman, Phys. Rev. B{\bf 33}, 8896 (1986).\\
$^c$J. Zaanen and G. A. Sawatzky, J. Solid State Chem. {\bf 88}, 8 (1990).\\
$^d$A. E. Bocquet {\it et al.}, Phys. Rev. B{\bf 46}, 3771 (1992).
\end{table}

First, applying a potential shift $w_d$ equally to all $d$ suborbitals
analogously to LMTO treatments, the derived value of $U$ is shown in Table I.
Comparison is provided with values obtained by the method of AZA, and
it is seen that the values obtained are 40-65\% of the
values obtained by AZA.  That our values are smaller is no surprise,
since our approach (of not disconnecting $d$ orbitals from other
orbitals) naturally allows additional screening to occur, by including
hybridization between $d$ orbitals and neighboring oxygen
$p$ orbitals.  Moreover, charge rearrangement between $e_g$ and $t_{2g}$
subshells reveals that there is some intra-$d$-shell screening in
the current approach.  In addition, our definition of the $d$ orbital
is not identical with that of AZA.

% FIG. 1
\begin{figure}[tbp]
\epsfysize=8cm\centerline{\epsffile{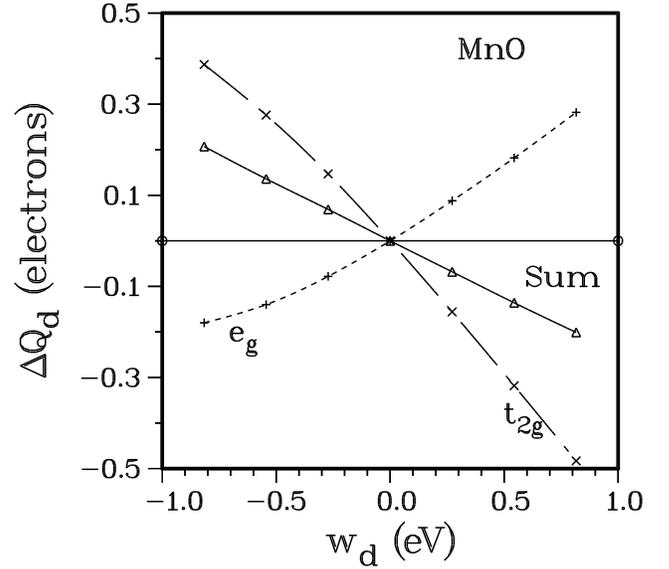}}
\caption{Change in on-site $d$ charge (solid line) in MnO resulting from a
potential shift $w_d$ applied to all $d$ states on a single Mn atom in
a four molecule supercell.  The total charge is decomposed into its
$e_g$ (short dashed line) and $t_{2g}$ (long dashed line) components.
\label{Fig1}}
\end{figure}

% FIG. 2
\begin{figure}[tbp]
\epsfysize=8cm\centerline{\epsffile{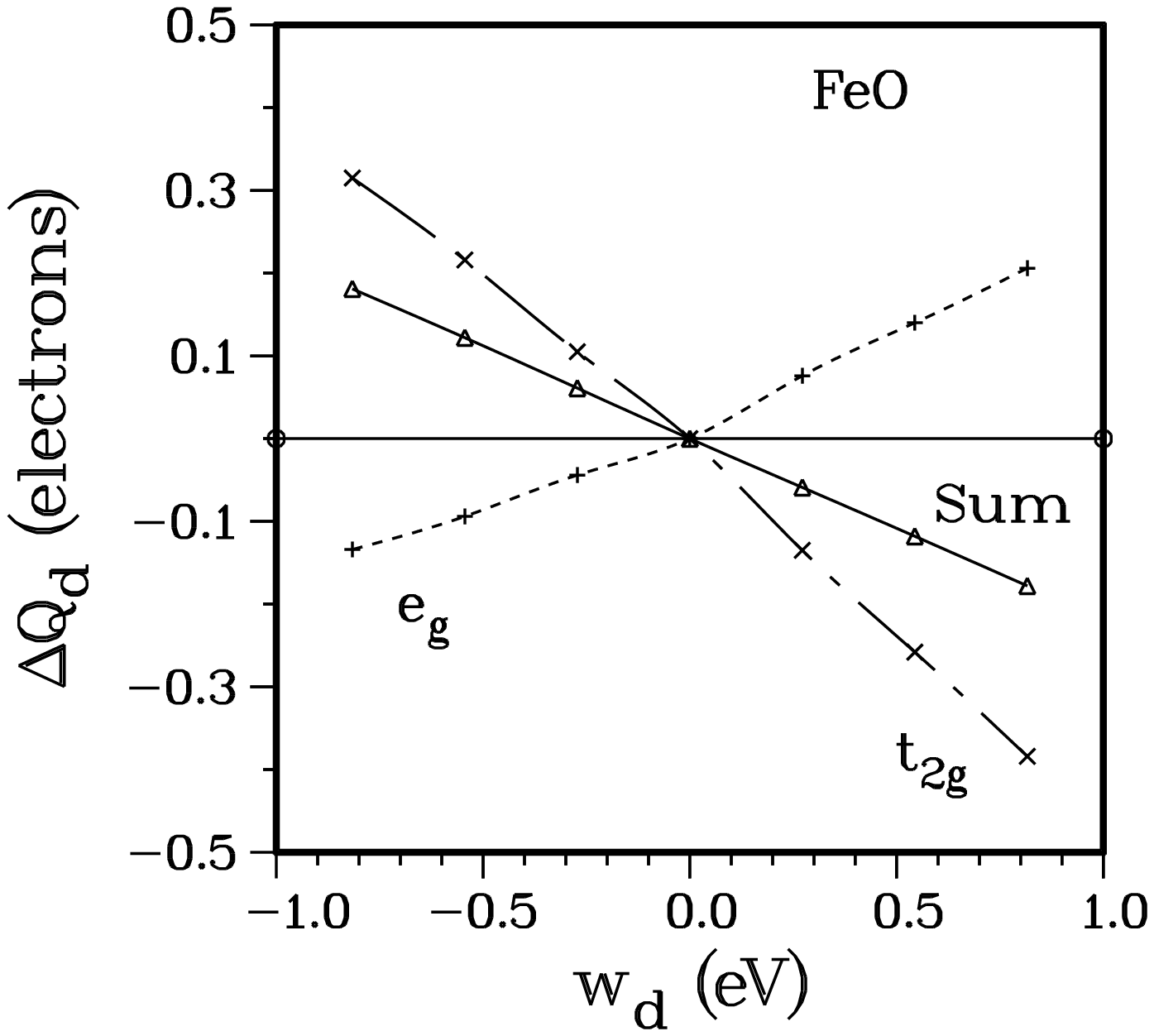}}
\caption{Change in $d$ charge in FeO, plotted as in Fig. 1.
\label{Fig2}}
\end{figure}

% FIG. 3
\begin{figure}[tbp]
\epsfysize=8cm\centerline{\epsffile{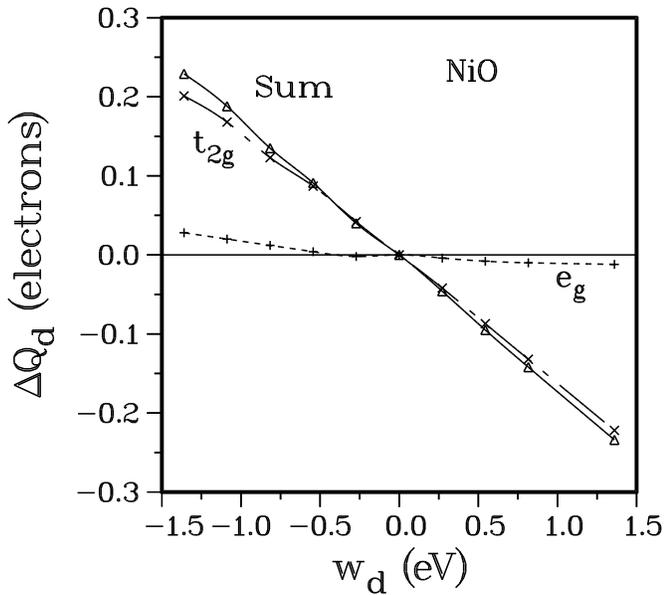}}
\caption{Change in $d$ charge in NiO, plotted as in Fig. 1.
\label{Fig3}}
\end{figure}

Figs. (1)-(3) for MnO, FeO, and NiO illustrate
the change in subshell charge with $w_d$ as well as the total change,
which is what determines $U$.  Taking MnO, for example, it is seen in
Fig. 1 that the effect of positive $w_d$ is to decrease $n_{t_{2g}}$
as expected, but $n_{e_g}$ instead {\it increases}.  Clearly charge
rearrangement within the $d$ shell is leading to a reduction in $U$
({\it i.e.} additional screening).  Similar behavior occurs for FeO
(Fig. (2)) while for NiO the $e_g$ charge remains almost unchanged as $w_d$
is varied.  Note
that, besides the differences in approach and in basis sets, the values
obtained in the AZA approach are evaluated for differences in $d$ charge
of unity.  Empirically determined values (obtained by comparing to
excited state data) lie in nearly all cases between our values and those of AZA.

The values of $U$ in Table I
are obtained as the first derivative of polynomial fits to the
$w_d$ vs.~$Q_d$ curve [Eq.~(\ref{eq:wvsQ})].
Figs. (1)-(3) indicate the $\Delta Q_d~vs.~w_d$
curve for shifts $w_d$ up to $\pm$ 0.8 eV for MnO, FeO,
and NiO.  The change in total $d$ charge is linear up to this size
of shift ($\approx$20-30\% of $U$).  Even for this size shift, however,
the individual $t_{2g}$ and $e_g$ contributions are beginning to
become nonlinear, as seen most clearly for MnO in Fig.~1.

\begin{table}[tbp]
\caption{$d$ shell charges, according to various definitions, for
transition metal monoxides from VO to NiO.}
\begin{tabular}{cccccc}
  Type       &  VO  &  MnO &  FeO &  CoO &  NiO  \\
\tableline
 Formal      &  3   &  5   &  6   &  7   &  8    \\
 On-site     &  3.67&  5.45&  6.22&  7.41&  8.22 \\
 Mulliken    &  3.09&  5.48&  6.44&  7.20&  8.40 \\
\end{tabular}
\label{table2}
\end{table}

On-site charges and Mulliken charges within LDA, for our basis
set, are compared in Table II.  The charges are less ionic than their
formal (dipositive) charge, as experience would suggest.  (Although
atomic charge within a crystal cannot be defined uniquely, it is 
widely accepted that `effective' ionic charges are nearly always
reduced by hybridization from their formal, full ionic values.)
Although VO is somewhat of an exception, the Mulliken charge does not 
differ more than 4\% from the on-site charge for these examples.
The response of Mulliken and on-site charges are very different,
however, with Mulliken charges varying more slowly.  
If one uses Mulliken charges rather than on-site charges
to obtain $U$, the resulting values are much larger:
3.8 eV for VO, 6.2 eV for MnO, and 11.1 eV for NiO.

% FIG. 4
\begin{figure}[tbp]
\epsfysize=14cm\centerline{\epsffile{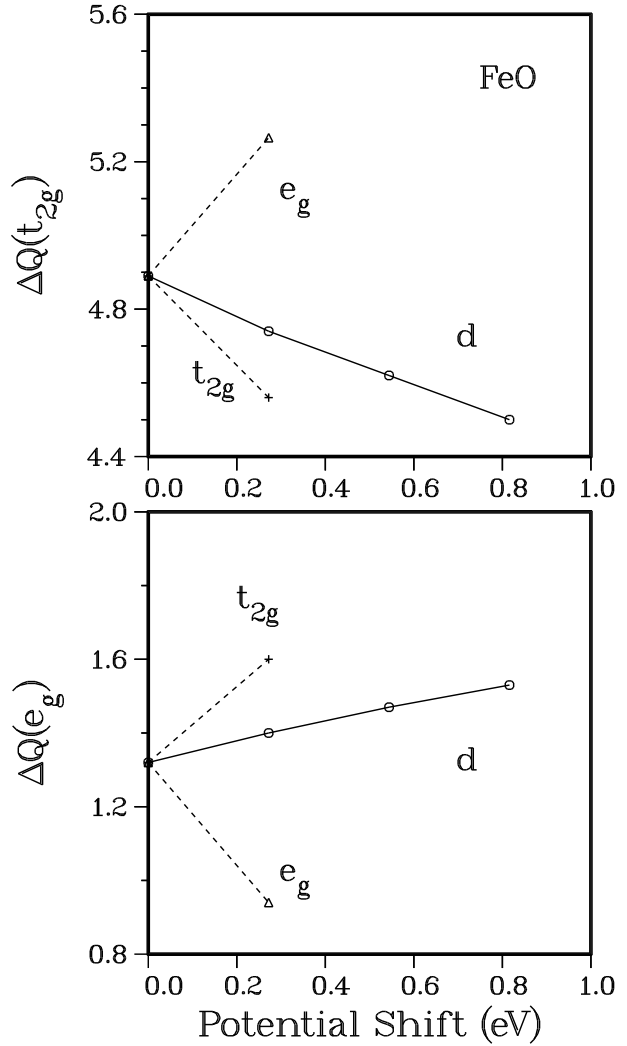}}
\caption{Change in subshell ($e_g$ and $t_{2g}$) charge in FeO,
resulting from a potential shift of only one of the subshells.
Top panel: change in $t_{2g}$ charge; bottom panel: change in $e_g$
charge.  The label indicates the type of applied potential shift $w$:
$d$ indicates shifts of all $d$ states; $t_{2g}$ ($e_g$) indicates
shift of only $t_{2g}$ ($e_g$) states.  Only positive energy shifts
are shown.
\label{Fig4}}
\end{figure}

\subsection{Sub-orbital Dependent $U$}
We have studied FeO charge redistribution when $e_g$ and $t_{2g}$
subshells are treated separately.  In Fig.~4 we present the change in
subshell charge when a shift in potential is applied individually
to the subshells.  In both cases, charge forced out of one subshell
by an upward shift in potential goes primarily into the other
subshell, amounting to very strong intra-$d$-shell screening in these
cases. Using Eq.~(20), 
we obtain, in eV$^{-1}$,
\begin{eqnarray}
\left(\matrix{U^{-1}_{t_{2g},t_{2g}} & U^{-1}_{t_{2g},e_{g}}\cr
              U^{-1}_{e_{g},t_{2g}} & U^{-1}_{e_{g},e_{g}}\cr}\right)
 = \left(\matrix{1.18 & -1.00 \cr
                -1.39 &  1.41\cr}\right),
\end{eqnarray}
which satisfies the sum rule of Eq.~(21).  The inverse is, in eV,
\begin{eqnarray}
\left(\matrix{U_{t_{2g},t_{2g}} & U_{t_{2g},e_{g}}\cr
              U_{e_{g},t_{2g}} & U_{e_{g},e_{g}}\cr}\right)
 = \left(\matrix{5.15 & 3.65 \cr
                 5.08 & 4.31 \cr}\right).
\end{eqnarray}
Recall that $U_{dd}$=4.6 eV, so in the usual orbital-independent   
treatment the corresponding matrix would be
\begin{eqnarray}
\left(\matrix{U_{t_{2g},t_{2g}} & U_{t_{2g},e_{g}}\cr
              U_{e_{g},t_{2g}} & U_{e_{g},e_{g}}\cr}\right)
 = \left(\matrix{4.6  & 4.6  \cr
                 4.6  & 4.6  \cr}\right).
\end{eqnarray}
Thus the behavior that looks rather peculiar in Fig. 4, and the negative
off-diagonal elements in Eq.(22), do
not lead to pathological behavior in the direct matrix ${\bf U}$.

% FIG. 5
\begin{figure}[tbp]
\epsfysize=7cm\centerline{\epsffile{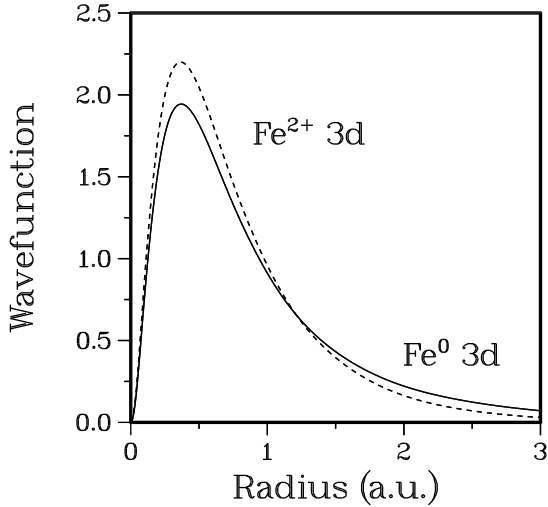}}
\caption{Comparison of the 3$d$
radial functions used in the LCAO
basis set as the Fe 3$d$ orbital to calculate the two values of $U$
reported in the text.  The functions are radial wave functions from neutral
and doubly ionized Fe atomic calculations, fit to a set of
eleven Gaussians.
\label{Fig5}}
\end{figure}

\subsection{Dependence of Local Orbital Shape}
The LDA+$U$ procedure is built around some choice of local orbital.
In an LCAO basis, this orbital is specified at the beginning, and
we have used neutral atom $d$ orbitals from atomic LDA calculations.
Another possible choice might be, say, the $d$ orbital from a positive
ion.  For FeO we have checked the effect of using the Fe$^{2+}$ $d$
orbital obtained from an atomic calculation on an isolated ion.  
The difference in radial density is shown in Fig. 5.
For the FeO (paramagnetic) solid,
the on-site charge of 6.22 electrons (Table II) changes 
to the rather peculiar value of 4.85 electrons, and the calculated
value of $U$ increases from 4.6 eV to 7.8 eV.  The total energy, however,
changes only by +0.12 eV/FeO, which is a very modest change (adding $f$
functions in an LCAO or LMTO calculation can result even larger 
changes, which are unimportant for most purposes).  It is clear that
the choice of $d$ orbital can affect the calculated value of $U$, 
certainly in the LCAO method but most likely in any calculational approach.

\section{Discussion}
The results presented in the previous section reflect a strong 
difference in response of the $e_g$ and $t_{2g}$ electrons, at
least to potentials of moderate strength.  Such
differences have been noted several times in the literature.  In
the context of the LDA+$U$ method, Solovyev {\it et al.} \cite{solovyev2}
have advocated using using separate values of $U$ for the two subshells
in perovskite structure transition metal oxides.  Their method of
obtaining $U_{\alpha}$ was a generalization of the standard method
described in Sec. II.  We, on the other hand, have adopted the differential
definition of $U$ that leads to a matrix $U_{\alpha \beta}$.

% FIG. 6
\begin{figure}
\epsfysize=7cm\centerline{\epsffile{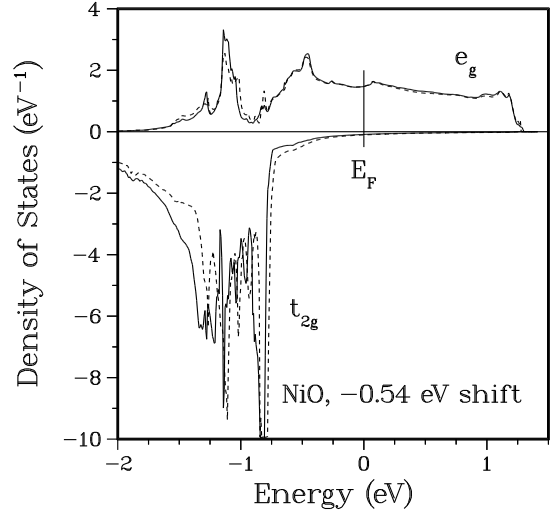}}
\caption{Ni $d$ density of states in NiO, before (dashed lines) and
after a shift of the $d$ states by -0.54 eV.  The $e_g$ ($t_{2g}$)
density is plotted upward (downward).  The Fermi levels have
been aligned.
\label{Fig6}}
\end{figure}

To begin to understand the response of the separate subshells, we show
in Fig. 6 the on-site $e_g$ and $t_{2g}$ densities of states (DOS) on a Ni
atom in an eight atom supercell of NiO, both before 
and after a downward shift of all $d$ states
by -0.54 eV.  The $t_{2g}$ states in the rocksalt structure
are weakly $dp\pi$ bonding and form a narrow band, whereas the $e_g$
states form $dp\sigma$ bonds that produce wider $e_g$ bands.  
From Fig. 3 it is seen that such shifts produce
negligible change in the on-site $e_g$ charge, with all the difference coming
entirely from the on-site $t_{2g}$ subshell.  This result is counterintuitive,
since the $t_{2g}$ DOS is full, and pulling it down seemingly cannot increase
its occupation.  The $e_g$ DOS is open shell and could accept charge,
but does not do so.

The resolution of this paradox lies in the change in the representation of
charge of the system by the LCAO basis functions as
a shift in potential is applied.  By looking at other 
local DOS, for both on-site and Mulliken charge decompositions,
we have found that a downward shift of $d$ states, which 
changes the degree and character of hybridization as well as the
probability of occupation, results in a
more active participation of the virtual orbitals in representing
the charge density.  To some extent this is a real effect: $d-p$
hybridization increases as the $d$ states are pulled down nearer
the $p$ bands, and what one calls the $d$ function, or the $d$
charge, becomes less well defined.  (The definition becomes clear
for well separated atoms, and perhaps for states well separated
in energy from any other states.)  The effect is present with 
other basis sets as well, but is more difficult to identify.

It is important to understand clearly the source of this paradox.
It arises because the ``$d$ charge density" is not a precisely and
objectively defined
quantity.  Although we are accustomed to thinking in terms of
five $d$ bands in a transition metal oxide that can be identified
and whose DOS integrates to five electrons per spin, this is a
fiction that becomes apparent as soon as orbital overlap becomes appreciable.
This difficulty has the same origin as the difficulty in defining
the ``$d$ orbital" to be used in the LDA+$U$ method.
These ambiguities are problems that must be lived with until a
better prescription can be formulated.

\section{Summary}
We have presented a reformulation of the method of obtaining $U$ for an
LDA+$U$ calculation.  The approach is based on a local orbital
expansion, which is a natural one considering that the $d$ orbital is
to be singled out and specified anyway.  We aim specifically to 
improve ground state properties rather than to account for 
spectroscopic data.

Values of $U$ using this approach are found to be only 40-65\% of the
values of Anisimov {\it et al.}  Most of this difference is understood
in terms of the definitions and procedures that are used in each case.
A novel feature here is the identification of an interaction matrix
that describes interactions that are non-diagonal in the suborbital
index, {\it e.g.} the change in energy of $t_{2g}$ states due to 
a change in $e_g$ charge.  The off-diagonal parts of this interaction
are expected to be strongly dependent on the environment of the ion,
and this expectation is borne out in our study of FeO.

There are important aspects of our approach that require further work.
The contribution to $U$ from the kinetic energy, and how it should be
dealt with, is one loose end.  The most appropriate choice of $d$
orbital is another question that may require some experience to
answer.  Carrying out LDA+$U$ studies to compare
with results using the previous LDA+$U$ method, and ascertaining the
effect of off-diagonal interactions, are however the main priority, and
this work is in progress.

\section{Acknowledgments}
We have benefited from discussions with V. I. Anisimov, G. E. Engel,
and I. I. Mazin.
E.C.E gratefully acknowledges support from the National Research 
Council - Naval Research
Laboratory Cooperative Postdoctoral Research Associate program.
This work was supported by the Office of Naval Research.
Calculations were carried out at the Arctic Region Supercomputer
Center and at the DoD Shared Resource Center at NAVO.

%
% references
%
 
%
\end{document}